# A simple interpretation of the growth of scientific/technological research impact leading to hype-type evolution curves

Marco Campani[1] and Ruggero Vaglio[2]

1) CNR Istituti SPIN, IOM e NANO, Corso Perrone 24, 16125, Genova (Italy)
2) Dipartimento di Fisica, Università di Napoli Federico II, Complesso Universitario di Monte S. Angelo, Via Cinthia 80126 Napoli (Italy) and CNR-SPIN

**Abstract**

The empirical and theoretical justification of Gartner "hype curves" is a very relevant open question in the field of Technological Life Cycle analysis. The scope of the present paper is to introduce a simple model describing the growth of scientific/technological research impact, in the specific case where science is the main source of a new idea driving a technological development, leading to "hype-type" evolution curves. The main idea of the model is that, in a first stage, the growth of the scientific interest of a new specific field (as can be measured by publication numbers) basically follows the classical "logistic" growth curve. At a second stage, starting at a later trigger time, the technological development based on that scientific idea (as can be measured by patent deposits) can be described as the integral (in a mathematical sense) of the first curve, since technology is based on the overall accumulated scientific knowledge.

The model is tested through a bibliometric analysis of the publication and patent deposit rate for Organic Light Emitting Diodes (OLED) scientific research and technology, as well as for other emerging technologies.

## 1. Introduction

Gartner "hype curves" (or "hype cycle") [1,2] provide a graphic representation of the evolution of visibility, maturity and adoption of specific technologies (and related products) as a function of time. The curves are aimed to provide an insight to company managers to follow its development and help making on time the right strategic decisions to optimize the company business in that field. The hype curves are generated by a superposition of a "bell shaped" curve describing the initial evolution of the new technology expectations and a classical s-shaped curve describing the technology maturity level [2,3]. The resulting curve presents a characteristic shape: an initial peak (technology trigger, peak of inflated expectations), followed by a dip (trough of disillusionment) evolving finally to a plateau (slope of enlightenment, plateau of productivity). The

Gartner hype cycle model has gained a very wide popularity, especially toward corporate clients, and its success appears being so overpowering to self-justify its validity. In contrast, the existing scientific literature on the subject has been still scarcely successful in finding experimental verifications of the Gartner hype cycle, as well as to furnish a valid theoretical frame. However interesting attempts towards empirical measurements of the hype of new technologies have been reported in the literature, using news-media and bibliometric indicators [4,5] and, more recently, web search traffic analysis [6].

The scope of the present paper is to contribute to furnish a theoretical basis for the hype curves. In particular, we will show that curves similar to the Gartner hype curves ("hype type" curves) can be generated by a very simple mathematical model that describes the evolution of the impact (hype), in the specific case where science is the main source of a new idea driving a technological development. In this simple frame, we can safely assume that the impact evolution is effectively measured by the use of bibliometrics (both basic research publications and patents) [7].

The mathematical model introduced in the following section is based on the idea that, following Gartner original approach [2,3], the overall hype curve shape is determined by the superposition of two simple curves describing two different phases of the scientific/technological growth process. For the first we will use the well known "logistic" curve, historically derived to describe the population growth/contagion processes. This curve shows an initial exponential growth of the impact (hype) of a new, appealing scientific research field, followed by a gradual tendency to reach a constant value and finally by a rapid (again exponential) decrease of the growth rate due to a natural saturation of the field. In this phase we can assume that the impact can be well measured by the basic research publication rate [7].

The second curve, starting at a later trigger time, describes the development of the impact of technological applications based on that scientific idea. The main assumption here, that is the core idea behind our work, is that technological developments are based on the overall accumulated scientific knowledge, so that the "technology growth curve" can be well described simply as the integral (in mathematical sense) of the "scientific growth curve". In this phase we will assume that the impact of the technology can be measured by the patent deposition rate [7].

In the third section, through the use of bibliometrics, as discussed above, we will show test data on Organic Light Emitting Diodes/Devices or OLED research and technology (over many conducted) indicating the validity of our approach, at least at a qualitative level.

**2. The hype-type model**

It is well known in the scientific community, at least at an intuitive level, that the emergence of a new scientific idea (or paradigm or concept) triggers, at an initial stage, an exponential rise of interest, whose impact (hype) can be measured using standard bibliometric indicators, as the number of published basic research (scientific and engineering) papers per year in that field. This exponential growth has the same origin of the population growth/contagion processes, and can be easily understood adapting epidemiology models to the dynamics of scientific interaction and it is very well experimentally verified for all high impact new emerging fields. This has been discussed, as examples, by Price [8,9] ,Crane [10] and many other authors [11,12].

With time this rise will gradually reduce and the overall interest (or impact, or hype) will eventually reach a maximum and than decay more or less symmetrically with time in respect to the "maximum impact" point. This decline process is again very similar to analogous processes observed in population growth/contagion cases and can be modeled mathematically in different ways, and at different levels of complexity (see, as examples, refs. [13,14]) .

The more simple model, that captures the essential features of the process, is the Logistic Growth mathematical model (or Verhulst model [15]). In this model it is assumed that the population growth rate is proportional to the product of the actual size of the population at the time t, times a term that accounts for the limited amount of resources (food, as an example) available to the population. This last term is proportional to the difference between the maximum resources limit (k) and the population size.

All this can be written mathematically as the following differential equation [15]:

$$\frac{dC(t)}{dt} = rC(t)\frac{[k - C(t)]}{k} \quad (1)$$

where $C(t)$ is the population at the time t, the proportionality constant $r$ is "the growth rate" and the constant $k$ is also called the "carrying capacity".
The solution can be easily found and can be written as:

$$C(t) = \frac{k}{1 + e^{-r(t-t_o)}} \quad (2)$$

that is called the logistic curve ($t_o$ is determined by the so called "initial conditions").

If we now identify in our case the "population" as the cumulative number of published basic research papers in a specific field (that is, of course, a good indicator of the scientific impact or hype of a specific field), we can derive the number of published papers per unit time as:

$$N(t) = \frac{dC(t)}{dt} = \frac{kre^{-r(t-t_o)}}{[1 + e^{-r(t-t_o)}]^2} \quad (3)$$

The logistic model well describes the initial exponential growth rate of the literature, generally observed, as already mentioned, in all cases of rapidly growing scientific fields and the following tendency to a decrease of the publication rate. The use of this model has been often criticized for its characteristic feature of being symmetric around the maximum increasing cumulative publication rate time [16]. Indeed in most cases reported in the literature, as well as in our own observations (see next section), the publication rate curve is not really symmetric around the maximum impact point. The decay rate indeed appears generally being somewhat larger in respect to the growth rate due to complex underlining phenomena, not included in the simple logistic model. Nevertheless it is generally recognized in the literature [13,14,16] that the model captures the essential features of the process of impact evolution in the purely "basic research phase", as can be measured by the scientific/engineering publication rate, and this justifies the use of eq. 3 in the following to describe this phase.

At a certain time after the basic research phase trigger time, technological developments start to appear (when the scientific idea contains in fact the potentiality for this to happen). The core idea behind our work is that the capability to develop new technological ideas depends on the overall accumulated scientific knowledge, so that the "technology growth curve" can be well described simply as the integral (in mathematical sense) of the "scientific growth curve". This idea can appear as too simple (or too "idealistic" in some sense) but certainly contains some truth : Indeed any advance in technology relies on the detailed knowledge of the fundamental properties and characteristics at the base of the exploited technology, properties and characteristics that are determined in the scientific/engineering phases.

Its justification will be reinforced "a posteriori", just by its capability to explain, at least in the considered specific case, the origin of the well assessed Gartner hype curve shape.

On the basis of this assumption we propose that, in the cases where a specific scientific idea generates a useful technology, the impact (hype) of the technology is given by the function :

$$P(t') = p\int N(t')dt = pC(t') = \frac{pk}{1+e^{-r(t'-t_o)}} \qquad (4)$$

where p is a proportionality constant and $t'= t-t^*$, $t^*$ being a parameter that regulates the delay from the trigger time of the scientific idea and the trigger time for technological developments, a significant delay time being naturally expected since practical application require some additional initial time to be generated.

The function $P(t')$ should be well measured by the number of patents deposited per unit time . In fact, though often criticized as an indicator for technological impact [17], patents have been shown to correlate well with the effective technological development [7, 18], so that patent statistics is a good indicator to be used in this frame.

In our model the overall evolution trend of the impact (hype) of a specific scientific/technological topic can be described using Eq. 3 for the "basic research phase growth curve" and Eq. 4 for the "technological phase growth curve"

Our "hype type" curve $H(t)$ is just simply the superposition (sum) of the curves describing the two phases :

$$H(t) = N(t) + P(t') \qquad (5)$$

with N(t) and P(t') given respectively by Eq. 3 and 4 .

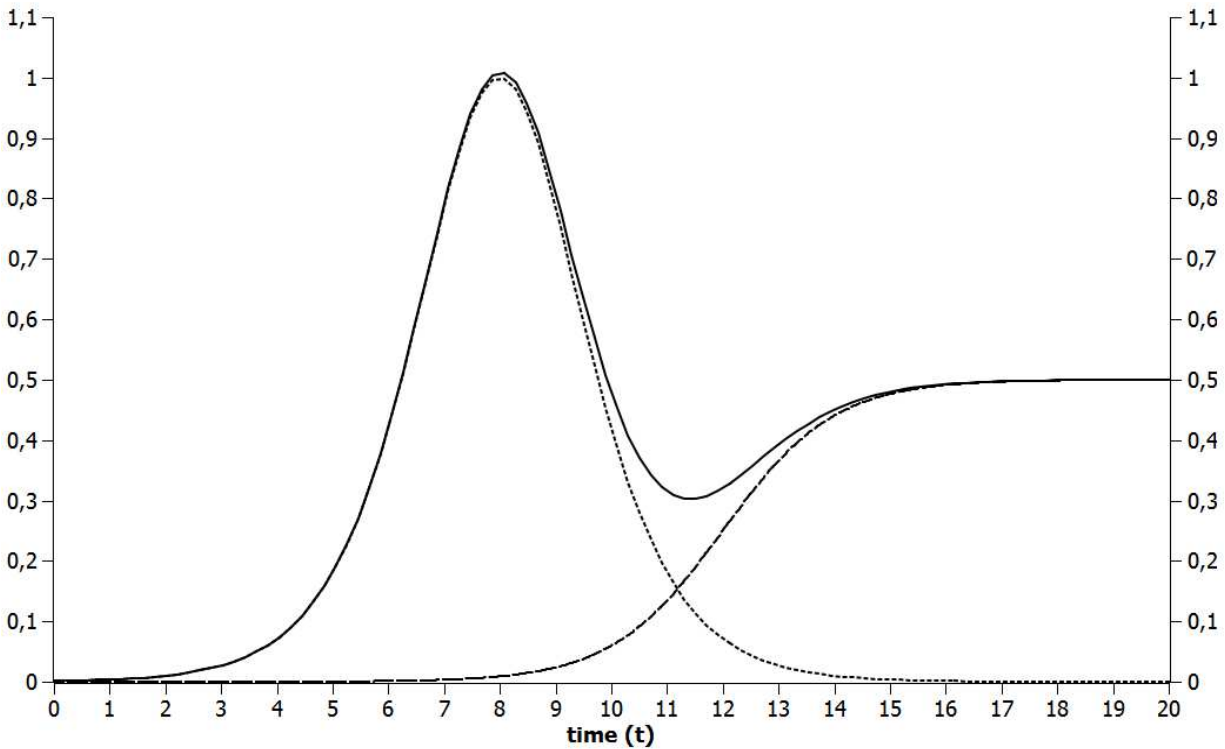

**Fig. 1.** Hype-type curve generated by Eq, 5 (full curve). Dotted and dashed lines represent, respectively, the impact of the basic research phase (rate of scientific/engineering papers publication, Eq. 3) and of the technological phase (rate of deposited patents, Eq. 4) respectively. On the horizontal axis is the time *t*, the parameters are *r=1, $t_o$=8, t\* = 4, k* and *p* are selected to normalize the dotted curve to *1* and the dashed curve to *0.5*.

Since the two curves measure different objects (publications and patents) a direct comparison of the maximum amplitude (hype) is not possible [3] . Therefore, as in the Gartner approach [1,2] in Fig. 1, Eqs. 3 and 4 are plotted selecting the parameters to normalize to 1 the maximum value of impact in the basic research phase and to 0.5 the maximum value of impact for the technological phase. The full continuous curve in Fig. 1 is simply Eq. 5, i.e. the superposition (sum) of the dotted and dashed curves introduced above, describing the evolution of the overall impact of basic research + technology. The overall curve shape, as already anticipated, is very similar the Gartner hype curves [1,2], also indicating a possible generality of the underlying processes.

It is worth underlining here that our model uses in fact the "Technological Life Cycle" (TLC) indicators measuring the impact of fundamental+applied research by the scientific and engineering papers publication rate and the following technological development phase by the patent deposit rate [19,20,21]. Our analysis is limited to the early technology growth stage so that the further application/production/ maturation phases are not considered here.

Of course the overall hype-type curve can differ in shape from that reported in Fig. 1 , assuming different values for the ratio of the basic research impact and technological impact, the time delay and the "growth rate" of the two distributions. In particular the dip in the curve is not observed for $t^* << t_o$.

**3. Model testing: the OLED case.**

Organic Light Emitting Diodes/Devices (OLED) is a relevant field of scientific and technological research. Basic research in the field of organic semiconductors has in fact lead to the production of high quality, reliable materials, and the technological development is leading to the design of complex device architectures with the target of producing low cost, high performance devices for a variety of applications (flexible and curved displays, transparent displays, large area lightning panels, intelligent windows, etc.).

To test the model described in the previous section we considered the OLED case, based on the following characteristics:
- this is certainly a case where science is the source of the ideas leading to the corresponding technological development;
- the scientific impact can be easily measured through bibliometric indicators having a very reduced ambiguity in search keywords;
- trigger time is 1990, in the full maturity of the "bibliometric age", so that spurious effects due the rapid increase of the overall yearly scientific publication rate is somewhat limited.

We have adopted the following methodology to extract the data:
- Scientific impact phase (basic + applied research) → bibliometric search on Thomson Web of Science Core collection and Scopus Elsevier database using the following keywords in the title: OLED/ORGANIC LED/ORGANIC LIGHT EMITTING. The search has reported 8179 results for ISI Web of Science and 8955 results for SCOPUS.
- Technological impact phase → patents from the following databases:
    Patentscope (WIPO - World International Patent Organization)         19614 results
    Espacenet (Free database of EPO - European Patent Office)            22928 results
    Acclaimip (FreePatentsOnline)                                        22993 results

The results obtained using different data bases where in all cases very similar, confirming data validity and reliability. The results are reported in Fig. 2, were the different search results using different database sources have been averaged out.

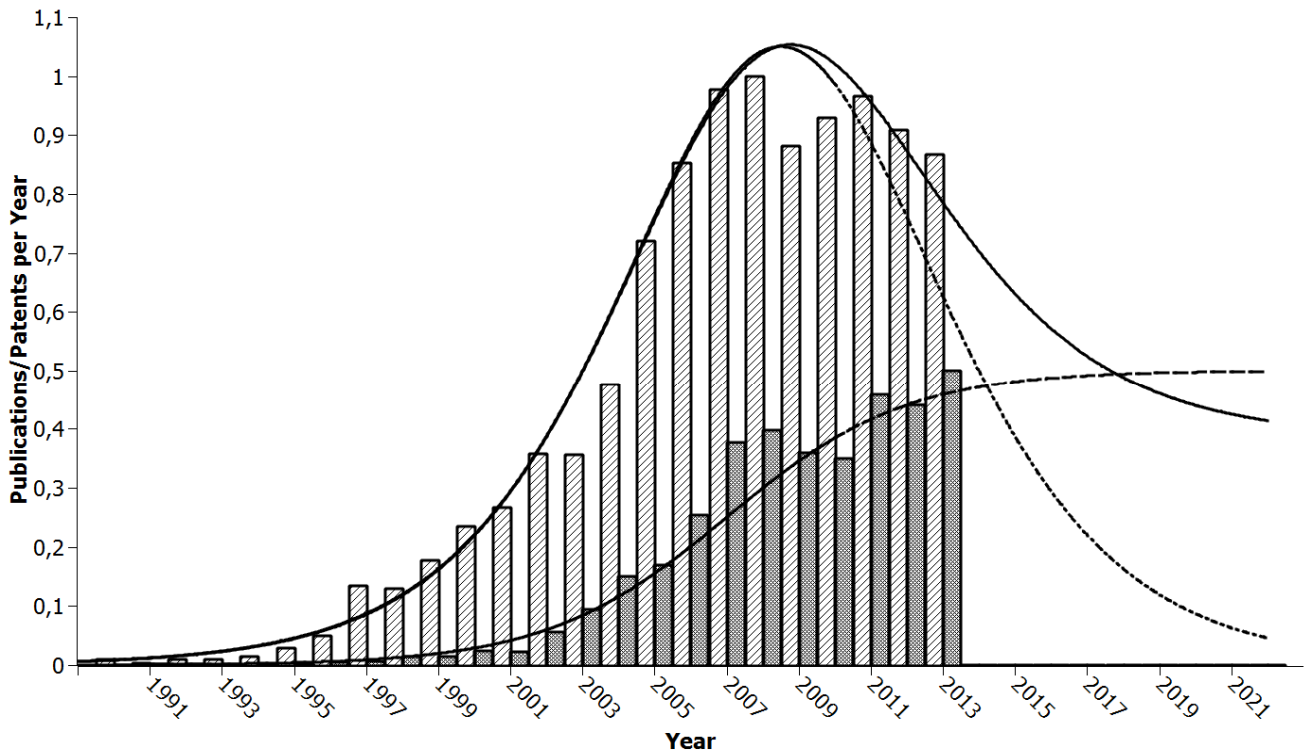

**Fig. 2.** Number of OLED scientific publications per year (clear gray histogram) and number of OLED deposited patents per year (dark gray histogram). Data are normalized stetting to 1 the maximum yearly publications and to 0.5 the maximum yearly patent deposits. The dotted curve represents Eq. 3, the dashed curve Eq. 4, the parameters have been chosen to give the best data description. The full continuous curve is the sum of the two dotted-dashed curves (Eq. 5).

The clear gray histogram represents the number of publications (scientific and engineering) per year (fundamental research impact) whereas the dark gray histogram represents the number of patents per year (technological impact). Data are normalized setting to 1 the maximum yearly publications and to 0.5 the maximum yearly patent deposits. The dotted and dashed curves are calculated by Eqs 3-4, to describe at the best the data. The full, asymmetric, continuous curve represents Eq. 5 .

The overall agreement , at least in terms of general trends, with our model is reasonable.

The difference in trigger times for research (1990) and technological applications (1995) is pretty clear in the figure. However, due to the small value of $t^*$ the characteristic dip of the hype curve is not observed here. Also the presence of a peak in publications and a saturation in patents is clearly exhibited, confirming general trends well clear from the current literature. In particular the s-shaped behavior with tendency to saturate to a plateau is very well verified for the patent deposit rate. The same behavior is clearly demonstrated in refs.[7,19].

The model predicts the future trends for OLED publications/patent data and implies a reduction of the publication rate over a factor two in respect to the maximum in 2020, and a constant rate of patents. The extension of the projection beyond few years (2020) would be meaningless, since at a certain stage the number of patents will clearly also decrease, and the impact of the technology has to be measured using different markers, based on production indicators, as in standard TLC models.

We have applied the same data analysis methodology used for OLED research to many other fields with analogous characteristics (C60, Giant Magnetoresistance, Superconducting $MgB_2$, and others). The obtained results where very similar to the OLED data, showing in all cases :
- a peak in the basic research publication rate
- a significant delay time (4-6 years) between the basic research and technology trigger times

- a tendency toward saturation of the patent deposit rate.

Analysis conducted on more recent scientific/technological research fields as Carbon Nanotubes or Graphene, has only shown the initial clear exponential increase of both rates of publications and patents numbers (with just an initial a tendency towards saturation for nanotubes) and the presence of a delay time between the two curves, but much more time will be necessary to check the full evolution curves of these emerging fields.

It is finally worth observing that the full continuous curve seems to describe the OLED publication data better than the dotted curve. In our opinion this could be due to the circumstance that Eq. 3 is intended to describe the evolution of the number of purely "basic" scientific/engineering papers. In fact, patents, especially if deposited by public scientific bodies, can produce a related "applied" research publication, whose number would be proportional to the number of patents. We can therefore further speculate that the full "hype type" curve deduced within our model can represent better than the dotted curve the overall publication rate, including basic (scientific+engineering) and applied research papers. To better elucidate this point a more deep analysis of the papers content would be required, that is outside the scope of the present work.

### 4. Conclusions

In the present paper we have proposed a new simple model for the evolution of the impact of technologies emerging from fundamental research, that well reproduces the characteristic "hype curves" introduced by Gartner [1,2]. The model describes the impact (or "hype") of technologies considering two different phases. In the first "basic research phase" the evolution of the impact is described by the use of the historically well set "logistic growth model", and can be measured through bibliometric methods, using the scientific/engineering publication rate. The second phase is the "technological phase" and the evolution of the impact can be measured, in this case, using the patent deposit rate. The truly innovative idea of our model is to assume that the capability to develop new technological ideas depends on the overall accumulated scientific knowledge, so that the "technology phase" curve can be described simply as the mathematical integral (starting at a later trigger time) of the logistic curve describing the scientific phase. The idea has a clear logical foundation, and appears to be "a posteriori" justified by the capability to reproduce the Gartner hype cycle and by its empirical success.

Our model uses indeed the classical TLC indicators, limiting the analysis to the fundamental+applied research phase and the following technological development phase, whereas the application/production/ maturation phases are not included in our study. The model has been successfully tested monitoring the evolution of the OLED scientific/ technological impact.

To conclude, since the scientific/tecnological research output is by itself a product, we suggest that the use of the "hype type" curves as described by our model can be of help both for Research Institutions and for individual researchers to make early strategic moves and to optimize their bibliometric indicators, in the same way that the Gartner hype cycle method helps Companies in their market strategies.

### Acknowledgments

The authors wish to thank Maria Grazia Maglione (ENEA) for sharing data on OLED market forecast and useful suggestions and Fabrizio De Filippis (University of Roma Tre) for careful reading of the manuscript.